\documentclass{PoS}

\def\tr{{\rm Tr}}

\def\bea{\begin{eqnarray}}
\def\eea{\end{eqnarray}}

\def\dash{\,\textendash\, }
\def\lmatrix{\left(\begin{array}}
\def\rmatrix{\end{array}\right)}

\def\msbar{\overline{\rm MS\kern-0.5pt}\kern0.5pt}

\title{Running coupling of the sextet composite Higgs model}

\ShortTitle{Running coupling of the sextet composite Higgs model}

\author{Zoltan Fodor\\
       University of Wuppertal, Department of Physics, Wuppertal D-42097, Germany\\
       Julich Supercomputing Center, Forschungszentrum J\"ulich, J\"ulich D-52425, Germany\\
       Eotvos University, Institute for Theoretical Physics, Budapest 1117, Hungary\\
       \email{fodor@bodri.elte.hu}}

\author{Kieran Holland\\
        University of the Pacific, 3601 Pacific Ave, Stockton CA 95211, USA\\
        Albert Einstein Center for Fundamental Physics, Bern University, Bern, Switzerland\\
        \email{kholland@pacific.edu}}

\author{Julius Kuti\\
        University of California, San Diego, 9500 Gilman Drive, La Jolla, CA 92093, USA\\
        \email{jkuti@ucsd.edu}}

\author{Santanu Mondal\\
        Eotvos University, Pazmany Peter setany 1, 1117 Budapest, Hungary\\
        MTA-ELTE Lendulet Lattice Gauge Theory Research Group, 1117 Budapest, Hungary\\
        \email{santanu@bodri.elte.hu}}

\author{\speaker{Daniel Nogradi}\\
        Eotvos University, Pazmany Peter setany 1, 1117 Budapest, Hungary\\
        MTA-ELTE Lendulet Lattice Gauge Theory Research Group, 1117 Budapest, Hungary\\
        \email{nogradi@bodri.elte.hu}}

\author{Chik Him Wong\\
        University of Wuppertal, Department of Physics, Wuppertal D-42097, Germany\\
        \email{cwong@uni-wuppertal.de}}

\abstract{The scale-dependent renormalized coupling of $SU(3)$ gauge theory coupled to $N_f = 2$ flavors of 
massless Dirac flavors in the sextet representation is presented in the range $0 < g^2 < 6.5$. This range includes
the location where the $\beta$-function turns zero in the $\msbar$ scheme to 3-loop and 4-loop approximations, however
our non-perturbative result shows a monotonically increasing $\beta$-function. Our lattice calclulations
are carried out at several lattice spacings allowing for a controlled continuum extrapolation. We also comment
on a recent similar calculation by Hasenfratz et al.}

\FullConference{The 33nd International Symposium on Lattice Field Theory - Lattice 2015,\\
		July 14-18, 2015\\
		Kobe International Conference Center, Kobe, Japan}

\begin{document}

\section{Introduction}
\label{introduction}

We continue our study of an $SU(3)$ gauge theory with $N_f = 2$ massless Dirac fermions in the sextet representation
\cite{Fodor:2012uu,Fodor:2012ty,Fodor:2012uw,Fodor:2012ni,Fodor:2014pqa,Fodor:2015vwa}. The primary motivation
for all of our studies is that this model may serve as 
the strongly coupled sector of electroweak symmetry breaking beyond the Standard Model.
In particular, the currently available numerical evidence is consistent with spontaneous
chiral symmetry breaking, generating exactly three Goldstone bosons before being eaten by three of the electroweak gauge 
bosons and with a light scalar particle to be interpreted as a composite Higgs boson.

In this contribution we address the renormalization group flow, more precisely its discrete variant, the step
scaling function. A renormalized coupling is defined in a finite volume gradient flow scheme and the step 
scaling function is calculated at several lattice volume pairs. These are then extrapolated to the continuum.
The result is a continuum discrete $\beta$-function corresponding to a finite change in the running 
scale $L \to sL$, in our case $s=3/2$, obtained from simulations on
$8^4 \to 12^4,\; 12^4 \to 18^4,\; 16^4 \to 24^4,\; 20^4 \to 30^4$ and $24^4 \to 36^4$. 
We pay particular attention to quantifying the size of systematic
uncertainties. In our setup the only source of these is the continuum extrapolation since we use rooted staggered
fermions and the mass can be set to zero exactly unlike in the Wilson fermion formulation. 

Since the result for the $\beta$-function is in the continuum it should be independent of the discretization
used as long as the same renormalization prescription \dash or scheme \dash is used for the coupling. Even
though the scheme used in \cite{Hasenfratz:2015ssa} is different from ours we do comment on the results found
there.

\section{Gradient flow scheme}
\label{flow}

\begin{figure}
\begin{center}
\includegraphics[width=7.5cm]{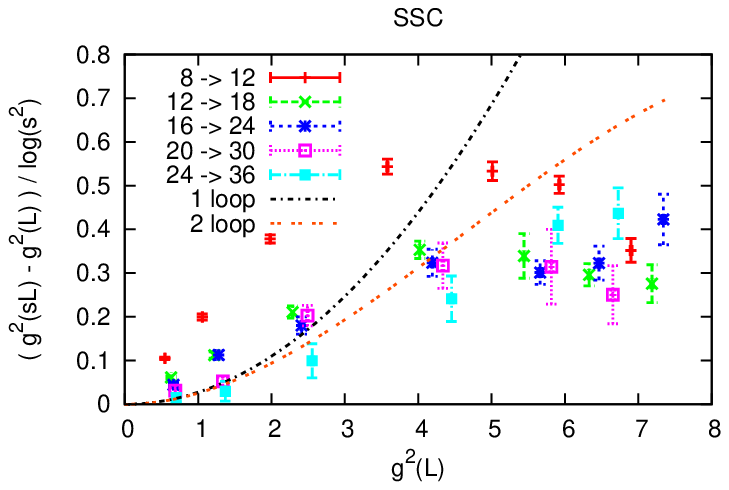} \includegraphics[width=7.5cm]{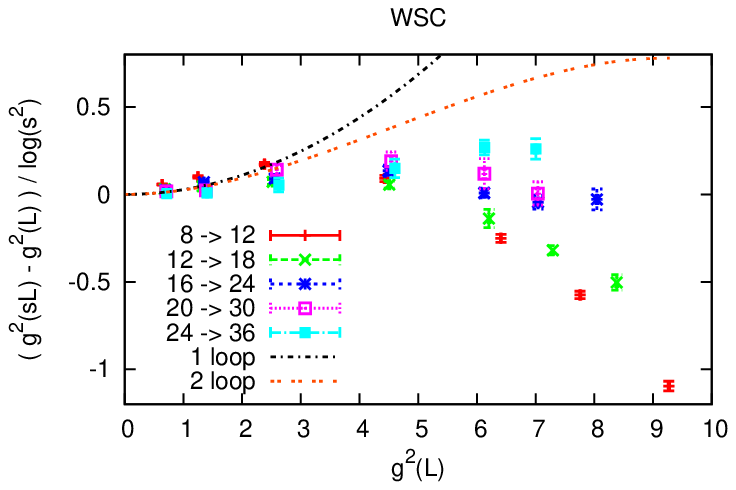}
\end{center}
\caption{Measured discrete $\beta$-function in the $SSC$ (left) and $WSC$ (right) discretizations; 
the data correspond to five sets of matched lattice volumes $L \rightarrow sL$ with $s = 3/2$.}
\label{betadata}
\end{figure}

In our work we use the renormalization prescription proposed in \cite{Fodor:2012td, Fodor:2012qh} for the
renormalized coupling. This scheme is based on the gradient flow  
\cite{Morningstar:2003gk, Narayanan:2006rf, Luscher:2009eq, Luscher:2010iy, Luscher:2010we,
Luscher:2011bx, Lohmayer:2011si}, is defined in finite 4-volume with periodic gauge fields and fermions 
which are anti-periodic in all 4 directions. More precisely, 
\bea
\label{g}
g_c^2 = \frac{128\pi^2\langle t^2 E(t) \rangle}{3(N^2-1)(1+\delta(c))}\;,\qquad E(t) = -\frac{1}{2} \tr F_{\mu\nu} F_{\mu\nu}(t)
\eea
where $N=3$ for $SU(3)$, $c = \sqrt{8t}/L$ is a constant that specifies the scheme and the factor
$\delta(c)$ can be found in \cite{Fodor:2012td, Fodor:2012qh}. In the present study we fix $c=7/20$.

\begin{figure}
\begin{center}
\includegraphics[width=7.0cm]{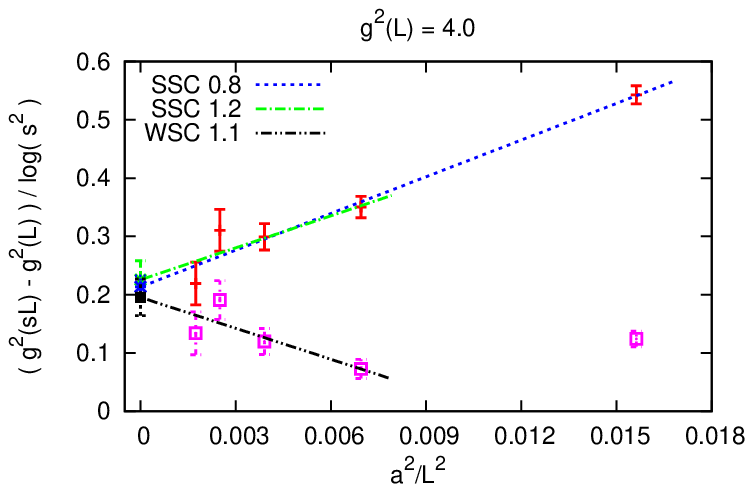} \includegraphics[width=7.0cm]{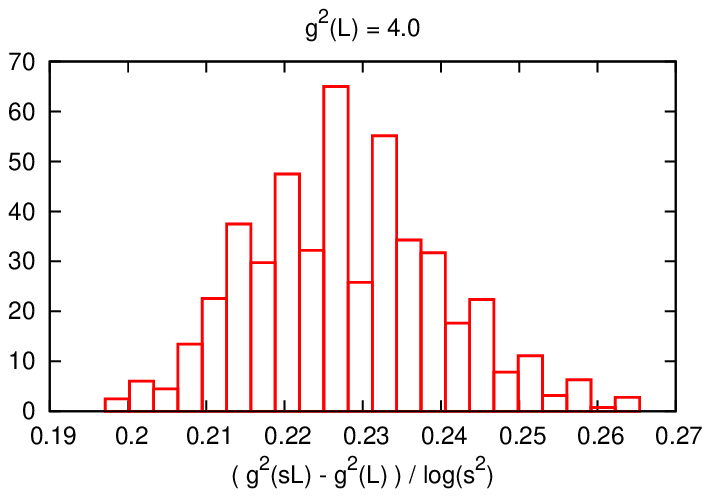} \\
\end{center}
\caption{Example of our continuum limit procedure for $g^2 = 4.0$ for $s=3/2$ and $c=7/20$. Right: 
the AIC-weighted histogram in the $SSC$ setup. Left: two representative examples
of the continuum extrapolations in the $SSC$ setup: one 5-point continuum extrapolation and one 4-point
continuum extrapolation. A representative example of 4-point continuum extrapolations in the $WSC$ is also
shown. In all cases the legend shows the $\chi^2/dof$ of the fits. See text for more details.}
\label{g4}
\end{figure}

So far the definition was given in the continuum and there are many ways to discretize it on the lattice
\cite{Fodor:2014cpa,Fodor:2014cxa,Ramos:2014kka,Ramos:2015baa}.
We use both Wilson plaquette and tree level improved Symanzik gauge action for the flow, 
tree level improved Symanzik gauge action for generating the configurations and the clover type 
discretization of $F_{\mu\nu}$ for measuring the observable $E$. These correspond to the $WSC$ and $SSC$
labels in the terminology of \cite{Fodor:2014cpa,Fodor:2014cxa}, respectively. 

The fermions are discretized using the rooted staggered formalism with 4 steps of stout-improvement
\cite{Morningstar:2003gk} and stout parameter $\varrho = 0.12$. The bare mass is set to zero. The applicability
of the rooting procedure in our finite volume setup at zero bare mass has been shown to hold in \cite{Fodor:2015zna}
to which we refer for more details.

The discrete $\beta$-function we seek to calculate is $( g^2(sL) - g^2(L) ) / \log( s^2 )$ as a function of 
the renormalized $g^2(L)$.
Clearly, in a lattice calculation with finite resources only a finite $g^2$-range can be covered for several reasons.

\section{Results}
\label{results}

\begin{figure}
\begin{center}
\includegraphics[width=8cm]{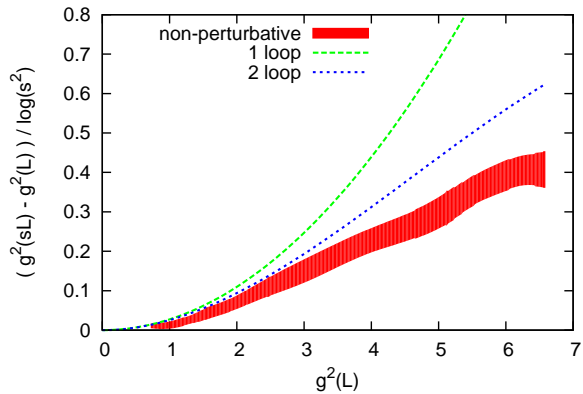} 
\end{center}
\caption{Continuum extrapolated discrete $\beta$-function for $s=3/2$ and $c=7/20$ using the $SSC$ setup.}
\label{final}
\end{figure}

In figure \ref{betadata} we show the measured discrete $\beta$-function values in both lattice discretizations $SSC$ and
$WSC$. There is a qualitative difference between the two cases at the finite lattice volumes shown: in the $WSC$
setup one observes a zero of the $\beta$-function at some finite lattice volumes, whereas in the $SSC$ setup
it stays positive for all lattice volumes. What happens at finite lattice volumes is of course
irrelevant, the only relevant question is how the continuum extrapolated result behaves.

The continuum extrapolation may be performed in the following manner, for more details see \cite{Fodor:2015zna}.
On each finite volume $L/a$ the dependence of $g^2(\beta)$ on the bare coupling $\beta$ is parametrized by polynomials
which will be used to interpolate to arbitrary $\beta$ values. A Kolmogorov-Smirnov test is applied to
various polynomial orders ($3,4,5$ on $L/a = 8, 12, 16, 18, 24$ and $3,4$ on $L/a = 20,30,36$) 
and only those combinations are deemed allowed which
lead to at least a 30\% Kolmogorov-Smirnov probability. All such allowed combinations are then used to
obtain continuum results at each $g^2$, assuming corrections are linear in $a^2/L^2$. The various continuum
results, corresponding to different allowed combinations of interpolations, are then binned into AIC-weighted
\cite{aic1, aic2, aic3}
histograms and the width of these histograms will be the estimate of the systematic 
uncertainty \cite{Durr:2008zz, Borsanyi:2014jba}. The statistical
uncertainty on the measured points of course lead to a statistical uncertainty on the interpolated polynomials
which in turn lead to a statistical uncertainty on the final result. The obtained statistical and systematic
uncertainties are then added in quadrature.
Figure \ref{g4} shows an example of our procedure at $g^2 = 4.0$. 

It was observed in \cite{Fodor:2015zna} that the $SSC$ discretization scales better than the $WSC$ variant
hence in the final result, figure \ref{final}, we show only the result corresponding to $SSC$. Both results however agree within
errors.

Clearly, the continuum $\beta$-function is monotonically increasing without a fixed point in the studied
range $0 < g^2 < 6.5$. We can of course not make any statement on the behavior for larger couplings than
our explored range, i.e. for $g^2 > 6.5$. However, in the $\msbar$ scheme the 3-loop and 4-loop
$\beta$-functions have a fixed point at $g^2 = 6.28$ and $g^2 = 5.73$, respectively and both are within our
available range. 
Some earlier lattice results  \cite{annatalks2, annatalks1, annatalks3} did in fact report consistency
with these perturbative results and hence consistency with an infrared fixed point.
The Schwinger-Dyson resummed perturbation theory is however consistent with our findings  
\cite{Sannino:2004qp, Ryttov:2010iz},
namely that approximation also predicts non-conformal behavior in the infrared.

\section{Comment on 1507.08260}
\label{comm}

In a recent work \cite{Hasenfratz:2015ssa} the discrete $\beta$-function was computed for the same model,
$SU(3)$ with $N_f = 2$ flavors of Dirac fermions in the sextet representation. The continuum scheme used
is similar, but not identical, to the one we used, despite the claim in \cite{Hasenfratz:2015ssa} that
they are identical.
In particular the coupling was defined from the gradient
flow with $c=7/20$ and $s=3/2$ for the step and the gauge fields were periodic, just like in our work.
However, the fermions were only 
anti-periodic in one direction and periodic in the rest, whereas in our work the fermions are anti-periodic
in all four directions.

Since the continuum schemes are {\em not} the same, quantitative agreement is not expected in the continuum extrapolated
results in general. In the perturbative region, for small $g^2$, agreement is nevertheless expected since
the $\beta$-function in both schemes should follow the universal 1-loop expression. The only expectation beyond
the perturbative region is that if one of the schemes shows a fixed point then the other scheme should show a fixed point
as well.

Nonetheless a quantitative comparison was made in \cite{Hasenfratz:2015ssa} between the two results and
a more than $3\sigma$ difference was noted for larger couplings at around $g^2 \simeq 5.0$. 
Let us reiterate, at larger coupling a quantitative comparison between the two continuum results is meaningless
since the two continuum schemes are not the same. The fact that different discretizations were used in the two
works (in \cite{Hasenfratz:2015ssa} Wilson fermions were used) 
is irrelevant since the continuum limit was taken in both. Note also that in earlier summaries
\cite{annatalks2, annatalks1, annatalks3} of the same work consistency with an infrared fixed point was reported.

\begin{figure}
\begin{center}
\includegraphics[width=8cm]{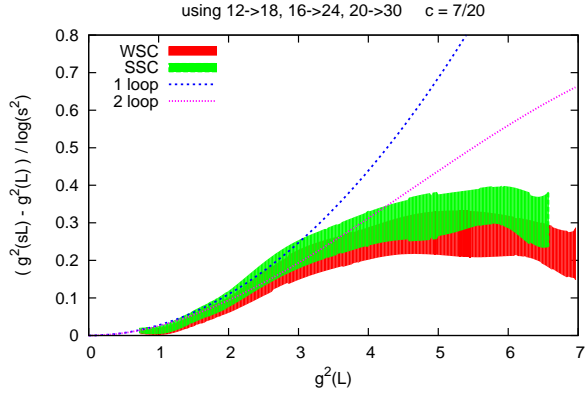} 
\end{center}
\caption{Our attempt to mimick the setup of \cite{Hasenfratz:2015ssa}. Only a subset of volumes are used,
$12^4 \to 18^4$, $16^4 \to 24^4$, $20^4 \to 30^4$, for the continuum extrapolation.}
\label{anna}
\end{figure}

Nevertheless it should be noted that the largest volume used in \cite{Hasenfratz:2015ssa} was $24^4$ (quoted
results on $28^4$ were not used in the continuum extrapolation) and the smallest was $12^4$. These corresponded
to the steps $12^4 \to 18^4$, $14^4 \to 21^4$ and $16^4 \to 24^4$.
In order to understand the effect of having smaller volumes and only 3 pairs, we repeated our continuum extrapolations using 
the steps $12^4 \to 18^4$, $16^4 \to 24^4$, $20^4 \to 30^4$ only. The procedure for the continuum extrapolation
and estimation of the systematic uncertainties is the same as in the
presented results in the previous section using all of our volumes. First the Kolmogorov-Smirnov test is applied to the 
interpolations on the subset of volumes
we use this time, then those which pass the Kolmogorov-Smirnov test are used for many different 
continuum extrapolations and the results are then binned into an AIC-weighted histogram. This procedure is repeated
for both the $WSC$ and $SSC$ discretizations. The obtained continuum result is shown in figure \ref{anna}.

Note that the $WSC$ and $SSC$ results agree within errors but the discrete $\beta$-function in the 
continuum is lower than our more reliable result in figure \ref{final} which used the full set of volumes;
the extrapolation excluding the largest volume is closer to the result in \cite{Hasenfratz:2015ssa}.
Note also that in our limited analysis using only $12^4, \ldots, 30^4$ we still have access to
a fine lattice spacing that is missing in \cite{Hasenfratz:2015ssa}, namely $20^4 \to 30^4$. 

It is important to observe that a major difference between the two works is that while in our formulation
we can set the bare fermion mass to zero, in the Wilson fermion formulation tuning is required. Table 1
in \cite{Hasenfratz:2015ssa} shows the estimated $\kappa_c$ values and the corresponding $m_c$ values. The difference
between $m_c$ and zero is in some cases 2, 3, 4 even 13 $\sigma$ and more. Even though not precisely tuned
$m_c$ might be less relevant for larger $\beta$ i.e. smaller physical volume which produces a larger gap in the
Dirac spectrum, one does observe 2 - 13 $\sigma$ deviations from zero at the small $\beta$ values as well. The
systematic uncertainty related to having not exactly massless fermions was not accounted for in
\cite{Hasenfratz:2015ssa}.

Hence we conclude that the most likely reason for a relatively low $\beta$-function in \cite{Hasenfratz:2015ssa}
is two-fold: one, the lattice volumes used were too small i.e. too large lattice spacings were used for
the continuum limit, and two, the systematic uncertainties were underestimated. The most probable source of
systematics is the tuning of the critical mass.

\section{Conclusion and outlook}
\label{conc}

We have presented recent results on the running coupling of $SU(3)$ gauge theory with $N_f = 2$ flavors of
massless Dirac fermions in the sextet representation. The goal is to study the model from as many angles as 
possible and to see whether the physics conclusions from the various approaches come together to form a coherent
conclusion about the infrared dynamics of the model or not. This is important because the model is clearly
quite different from QCD and is more difficult to study in the continuum limit in particular. Our previous
results were consistent with spontaneous chiral symmetry breaking which is also consistent with the present
study at least up to the maximal coupling $g^2 \simeq 6.5$ accessible to our numerical simulations.

\section*{Acknowledgments}

This work was partially supported by the DOE grant DE-SC0009919,
by the DFG grants SFB-TR 55 and by the NSF grants 0704171, 0970137 and 1318220,
and by the grant OTKA-NF-104034.
We received major support from an ALCC Award on the BG/Q Mira of ALCF.
Computations were also carried out at the GPU clusters of Fermilab,
the University of Wuppertal, University of California San Diego and Eotvos University using
the CUDA port of the code \cite{Egri:2006zm} and 
on Juqueen at FZJ. Szabolcs Borsanyi, Sandor Katz and Kalman Szabo are acknowledged for their help and code development.
KH wishes to thank the Institute for Theoretical Physics and the Albert
Einstein Center for Fundamental Physics at the University
of Bern and the Schweizerischer Nationalfonds for their support.

\end{document}